\documentclass[twocolumn,preprintnumbers,amsmath,amssymb]{revtex4}
\newcommand{\be}{\begin{equation}}
\newcommand{\ee}{\end{equation}}
\usepackage{graphicx}
\usepackage{multirow}

\begin{document}

\title{Calculation of Magnetic Field Noise from High-Permeability Magnetic Shields and Conducting Objects with Simple Geometry}

\author{S.-K.~Lee}
\email{lsk@princeton.edu}
\author{M.~V.~Romalis}
\affiliation{Physics Department, Princeton University, Princeton, NJ 08544, USA}
\date{\today}

\begin{abstract}
High-permeability magnetic shields generate magnetic field noise that can limit the sensitivity of modern precision measurements. We show that calculations based on the fluctuation-dissipation theorem allow quantitative evaluation of magnetic field noise, either from current or magnetization fluctuations, inside enclosures made of high-permeability materials. Explicit analytical formulas for the noise are derived for a few axially symmetric geometries, which are compared with results of numerical finite element analysis. Comparison is made between noises caused by current and magnetization fluctuations inside a high-permeability shield and also between current-fluctuation-induced noises inside magnetic and non-magnetic conducting shells. A simple model is suggested to predict power-law decay of noise spectra beyond quasi-static regime. Our results can be used to assess noise from existing shields and to guide design of new shields for precision measurements.
\end{abstract}

\maketitle

\section{Introduction}
Passive magnetic shields are frequently used in precision measurements to create a region in space which is magnetically isolated from the surroundings \cite{Mager1970}. A few layers of nested shells made of high-permeability metals, such as mu-metal, routinely provide in table-top experiments a quasi-static shielding factor in excess of $10^4$. Such a shield, on the other hand, generates thermal magnetic field noise which often exceeds the intrinsic noise of modern high-sensitivity detectors such as  superconducting quantum interference devices (SQUIDs) and high-density alkali atomic magnetometers \cite{Allred2002}.

 Magnetic field noise generated by thermal motion of electrons (Johnson noise current) in metals has been much studied in the past in the context of applications of SQUID magnetometers \cite{Varpula1984, Clem1987}, and more recently as a source of decoherence in atoms trapped near a metallic surface \cite{Henkel2005}. A majority of these works were devoted to low frequency noise from Johnson noise current in non-magnetic metals. A few authors also considered noise from high-permeability metals of flat geometry.  The calculations presented in these works, however, were not particularly amenable to extension to other geometries, such as those of cylindrical shields often used in table-top experiments. Nenonen et al., for example, used calculation of noise from an infinite slab to estimate noise inside a cubic magnetically shielded room for biomagnetic measurements \cite{Nenonen1996}. As shown below, the validity of such extrapolation is not immediately clear, given the image effect of high-permeability plates. Lack of explicit formulas and qualitative scaling relations for magnetic field noise from high-permeability shields have caused some confusion about the contribution of such noise in certain experiments (See discussions in Refs.~[\onlinecite{Munger2005,Budker2004}].).

Among different strategies that have been demonstrated to calculate magnetic field noise \cite{Varpula1984, Clem1987, Roth1998}, a particularly versatile method is the one based on the generalized Nyquist relation by Callen and Welton \cite{Callen1951, Clem1987}, which later led to the fluctuation-dissipation theorem. Here the noise from a dissipative material is obtained from calculation of power loss incurred in the material by a driving magnetic field. Sidles et al., for example, presented a comprehensive analysis of the spectrum of magnetic field noise from magnetic and non-magnetic infinite slabs with a finite thickness using this principle \cite{Sidles2003}.

A particularly useful feature of the power-loss based noise calculation is that it allows calculation of noise from multiple physical origins, including Johnson noise current in metals and domain fluctuations in magnetic materials. The noise of the latter kind in ferromagnets, which can be associated with magnetic hysteresis loss, was previously studied for toroidal transformer cores where field lines were confined in the core material \cite{Durin1993}. In a recent work \cite{Kornack2007} Kornack et al.  measured magnetic field noise in the interior of a ferrite enclosure with an atomic magnetometer which was consistent with predictions based on numerical calculation of power loss in the ferrite. The same paper also presented results of analytical calculations of the noise inside an infinitely long, high-permeability cylindrical tube.

In this work we show how similar calculations can be performed for other geometries with cylindrical symmetry, and derive a general relationship between magnetic field noises from current and magnetization fluctuations in shields with such geometries. For metallic shields, we show that the Johnson current-induced noise is either suppressed or amplified, depending on the shape of the shield, due to a high permeability. This partly explains previous confusion about noise contributed by magnetic metals. Analytical calculations leading to our key results were confirmed by numerical calculations on representative geometries using commercial finite element analysis software. In order to explain frequency dependence of noise from metallic and magnetic plates reported in literature, we propose a simple model which correctly predicts observed power-law decays in noise spectra. We also present in the Appendix analytical calculations of noise from non-magnetic conducting objects that can model other common experimental parts used in precision measurements.

\section{Principles}
The principle of calculating magnetic field noise from energy
dissipation in the source material has been demonstrated by several authors.
For example see Refs.~ [\onlinecite{Clem1987, Henkel2005, Sidles2003}]. The arguement is
summarized as follows. If at a point $\vec{r}$ there is a fluctuation of
magnetic field along direction $\hat{n}$, given by its power spectral
density $S_{B}\left( f\right) $, an N-turn pickup coil located at $\vec{r}$
directed along $\hat{n}$ will develop a fluctuating voltage, according to the
Faraday's law, with power spectral density
\be
S_{V}\left( f\right) =A^{2}N^{2}\omega ^{2}S_{B}\left( f\right) .
\label{Svf1}
\ee
Here $\omega =2\pi f$ and $A$ is the area of the pickup coil, assumed to be
small so that the field is uniform over the area. We further assume that the
coil is purely inductive, for example by making it superconducting, so that
in the absence of an external material (noise source) there is no voltage
fluctuation due to conventional Nyquist noise, $S_{V,coil}=4kTR_{coil}=0$. Now
assume that we take the pickup coil and the material responsible for the
noise as a single effective electronic element, whose small-excitation
response is characterized by an impedence $Z.$ The fluctuation-dissipation theorem applied to this system states that the voltage fluctuation at the terminals of the pickup
coil is related to the real part of $Z$, $\mathop{\rm Re}[Z(f)]\equiv R_{\mathrm{eff}}$, by
\be
S_{V}\left( f\right) =4kT\,%
R_{\mathrm{eff}}\left( f\right) .
\label{Svf2}
\ee
Here the system is assumed to be at thermal equilibrium at temperature $T$, $%
k$ is the Boltzmann constant, and the effective resistance $R_{\mathrm{eff}}$ is obtained from the time-averaged power dissipation in the system
\be
P\left( f\right) =\frac{1}{2}I^{2}R_{\mathrm{eff}}\left( f\right)
\label{Pf}
\ee
incurred by an oscillating current $I\left( t\right) =I\,\sin \omega t$
flowing in the pickup coil whose amplitude $I$ is small so that the response
is linear. In the absence of the resistance of the pickup coil itself, the
power dissipation is entirely due to the loss in the material driven
electromagnetically by the current $I\left( t\right) $. From Eqs.~(\ref{Svf1},\ref{Svf2},\ref{Pf}) this power determines the magnetic field noise by
\be
\delta B\left( f\right) \equiv \sqrt{S_{B}\left( f\right) }=\frac{\sqrt{4kT}%
\sqrt{2P\left( f\right) }}{ANI\omega }.
\label{Bfromp}
\ee

Since the power $P$ scales quadratically with the driving dipole $p\equiv ANI$
in the linear response regime, the above equation is independent of the size
and driving current of the pickup coil. The usefulness of this expression
lies in the fact that in most cases, calculation of power loss is much
easier than that of magnetic field noise, the latter requiring incoherent
sum of vectorial contributions from many fluctuation modes inside the source
material.

For high-permeability metals and ceramics used for magnetic shields the
primary sources of power loss at low frequencies (\mbox{$\lesssim$} 1 MHz) are eddy current loss $P_{\text{eddy}}=\int_{V}\frac{1}{2}\sigma
E^{2}dv$ and hysteresis loss $P_{\text{hyst}}=\int_{V}\frac{1}{2}\omega \mu
^{\prime \prime }H^{2}dv$ \cite{Lin2004}\cite{footnote1}. Here $\sigma $ is the conductivity, $\mu ^{\prime
\prime }$ is the imaginary part of the permeability $\mu =\mu ^{\prime
}-i\mu ^{\prime \prime }$, and the integrals are over the volume of the
material in which oscillating electric and magnetic fields of amplitude $E$
and $H$, respectively, are induced by $I(t)$.

For a given driving dipole strength $p$, the eddy current $j=\sigma E$ is
proportional to the frequency $\omega $, therefore $P_{\text{eddy }}$ leads
to a frequency independent (white) noise according to Eq.~(\ref{Bfromp}), to the extent
that $\sigma $ is frequency independent. On the other hand, $P_{\text{hyst}}$,
assuming frequency-independent $\mu $, leads to a noise with $1/f$ power
spectrum, which is indeed observed in experiments with ferromagnetic
transformer cores \cite{Durin1993}. In what follows we will denote the noises
associated with $P_{\text{eddy}}$ and $P_{\text{hyst}}$ by $\delta B_{\text{curr}}$ and $\delta
B_{\text{magn}}$, respectively.

\section{Power loss calculation for high-permeability shields with cylindrical symmetry}
In this section we calculate power dissipation in high-
permeability shields with cylindrical symmetry when the driving dipole is on
and along the axis of the shield. See Fig. 1(a) for a representative
geometry. We restrict ourselves to  quasi-static regime where the magnetic
field amplitude inside the shield material is given by  its dc value,
ignoring perturbation due to induced (eddy) currents which is proportional to the frequency. The power dissipation when the dipole is at other
locations and along other directions can be calculated numerically with, for instance, a
three-dimensional finite elememt analysis software commonly used for power
loss calculations in transformer cores.

\begin{figure}
\includegraphics[width=8cm]{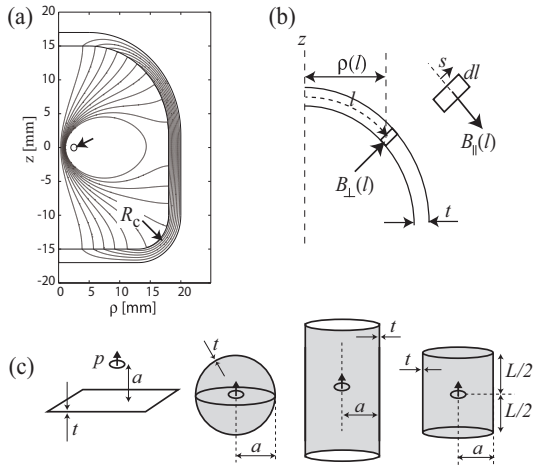}
\caption{(a) Magnetic field lines inside a high-permeability ($\mu_r$ = 1000) shield with cylindrical symmetry (mirror symmetry is not assumed). The field is generated by a current loop, carrying 1 A dc, whose cross section is indicated by a small circle. The eleven field lines enclose magnetic flux of $n\Delta\Phi, n = 1,2,\cdots,11,$ where $\Delta\Phi = 10^{-10}$ Wb. (b) Cross section of the shield in the $ \rho$-$z$ plane.  (c) Four geometries considered in the power loss calculation: infinite plate, sphere, infinite cylinder, and finite closed cylinder. The cylinders and the sphere are hollow shells of thickness $t$}
\label{figure1}
\end{figure}

Figure 1(a) also shows several magnetic field lines, calculated numerically,
in the $\rho $-$z$ plane around the shield generated by a current loop
modeling a driving dipole. Two features are noticeable. First, the field
lines entering the shield are very nearly normal to the surface,
reflecting the well-known boundary condition involving a high-permeability
material. Second, most of the field lines reaching the shield are
subsequently confined within the thickness of the shell, running
nearly parallel to the profile of the shield in the $\rho $-$z$ plane.
This, combined with the condition $\nabla \times \vec{B}=0$, requires that the field lines are nearly uniformly
spread within the thickness of the shield. For a shield
surface with radius of curvature (in the $\rho $-$z$ plane) $R_c$, it can be
shown that the variation of the field strength across the thickness $t$ of
the shield is $\delta B_{\parallel }/B_{\parallel }\approx t/R_c$, where $%
B_{\parallel }$ is the field component parallel to the shield in the $\rho $-$z$
plane. The condition for the field confinement can be estimated, from
dimensional consideration, to be $\mu _{r}t/a\gg 1$, where $\mu _{r\text{ }}$%
is the relative permeability, and $a$ is the characteristic distance between
the driving dipole and the shield surface \cite{footnote2}. Since the same factor $\mu _{r}t/a$ also determines
the shielding factor \cite{Mager1970}, we can assume this condition is satisfied if the shell is to function as a magnetic shield in the first place.
In summary, we assume the following for our calculations: (1) $\mu _{r}\gg 1$ so that the normal entrance boundary condition is satisfied. (2) $\mu _{r}t/a \gg 1$ so that most of the field lines, once entering the shield material, are confined within the thickness of the shield. (3) $t/R_c\ll 1$ for most part of the shield so that the confined field amplitude is uniform in the direction normal to the shield surface \cite{footnote3}.
\subsection{Eddy-current loss}

Suppose that the driving dipole is oscillating sinusoidally at a frequency $%
\omega $, $\vec{p}(t)=p\widehat{z}\sin \omega t$. We want to calculate, to the lowest order in $\omega $, the eddy current in the shield which is symmetric around the $z$-axis. We define the
position of an arbitrary point in the shield in the $\rho $-$z$ plane by
coordinate $(l,s)$ as shown in Fig. 1(b). Here $l$ defines a position in the midplane of the shield by measuring its distance from the $z$ axis along the cross-section of the shield. The coordinate $s$ represents the normal distance of a point from the midplane, $-t/2\leq s\leq t/2$.
Since we are interested in a thin-walled shell, we ignore the variation of the radial coordinate $\rho$ on $s$: $\rho(l,s)\approx\rho(l,s=0)\equiv\rho(l)$. Our assumptions in the preceding paragraph imply that the magnetic field within the shield material is parallel to the line defining the $l$ coordinate, and its amplitude $B_{\parallel}=B_{\parallel}(l)$ depends only on $l$. Finally we define $B_{\perp}(l)$ as the amplitude of the magnetic field entering the inner surface of the shield at $(l,s=-t/2)$.

In three dimensions, a point $(l,s)$ corresponds to a ring, and we define $%
\Phi (l,s)$ as the amplitude of the flux generated by the driving dipole $%
\vec{p}(t)$ that threads the ring. Then the amplitude of the eddy current
flowing along the ring is
\be
j_{\phi }=\sigma E_{\phi }=\sigma \cdot \omega \cdot \Phi (l,s)/2\pi \rho(l).
\label{jphi}
\ee
If all the field lines are confined within the shield, a ring on the outside surface of the shield has no net flux in it, $\Phi(l,s=t/2)=0$. For all other $s$, $\nabla\cdot\vec{B}=0$ dictates that
\be
\Phi (l,s) = 2\pi \rho (l)\left( \frac{t}{2}-s\right) \,B_{\parallel }(l).
\label{fluxring}
\ee

From Eqs.~(\ref{jphi},\ref{fluxring}) the eddy-current loss is
\begin{eqnarray*}
P_{\text{eddy}} &=&\int_{0}^{l_{\max }}\int_{-t/2}^{t/2}\frac{1}{2}\sigma E_{\phi
}^{2}(l,s)\,2\pi \rho (l)\,ds\,dl \\
&=&\int_{0}^{l_{\max }}\int_{-t/s}^{t/s}\frac{1}{2}\sigma \omega ^{2}\left(
\frac{t}{2}-s\right) ^{2}B_{\parallel }^{2}(l)\,2\pi \rho (l)\,ds\,dl \\
&=&\pi \sigma \omega ^{2}\int_{0}^{l_{\max }}B_{\parallel }^{2}(l)\,\rho
(l)\,dl\,\int_{-t/2}^{t/2}\left( \frac{t}{2}-s\right) ^{2}ds \\
&=&\frac{1}{3}\pi \sigma \omega ^{2}\,t\,\beta ,
\end{eqnarray*}
where the configuration integral $\beta $, having a dimension of flux
squared, is
\be
\beta =\int_{0}^{l_{\max }}t^{2}B_{\parallel }^{2}(l)\,\rho (l)\,dl.
\label{beta1}
\ee
This expression can be reduced to a form more useful in practical
calculations by expressing $B_{\parallel }(l)$ in terms of $B_{\perp }(l)$.
From $\nabla \cdot \vec{B}=0$ it follows that $tB_{\parallel }(l)\rho(l)=\int_{0}^{l}B_{\perp }(l^{\prime })\,\rho (l^{\prime })\,dl^{\prime }$. Therefore,
\be
\beta =\int_{0}^{l_{\max }}\left[ \int_{0}^{l}B_{\perp }(l^{\prime })\,\rho
(l^{\prime })\,dl^{\prime }\right] ^{2}\frac{1}{\rho (l)}dl.
\label{beta2}
\ee
\subsection{Hysteresis loss}
The hysteresis loss arises from a phase delay in the magnetic
response of a material to the applied oscillating magnetic field. For most
soft magnetic materials used for magnetic shields, this delay is small at
frequencies below $\sim $1 MHz. In the following we assume that the shield
has a constant permeability throughout its volume with $\mu ^{\prime \prime
}\ll \mu ^{\prime }\approx \mu _{r}\mu _{0}.$ The expression
for $P_{\text{hyst}}$, to the first order in $\mu ^{\prime \prime
}$, can then be obtained as follows.
\begin{eqnarray*}
P_{\text{hyst}} &=&\int_{V}\frac{1}{2}\omega \mu ^{\prime \prime }
H^2
\,dv \\
&=&\int_{0}^{l_{\max }}\int_{-t/2}^{t/2}\frac{1}{2}\omega \frac{\mu ^{\prime
\prime }}{\mu ^{\prime 2}}B_{\parallel }^{2}(l)\,2\pi \rho (l)\,ds\,dl \\
&=&\pi \omega \frac{\mu ^{\prime \prime }}{\mu ^{\prime 2}}\frac{1}{t}\beta .
\end{eqnarray*}
Therefore, both $P_{\text{eddy}}$ and $P_{\text{hyst}}$ are proportional to $%
\beta $. It follows that the ratio between magnetization- and current-
induced noises in a cylindrically symmetric shell measured on and along the
axis is
\be
\frac{\delta B_{\text{magn}}}{\delta B_{\text{curr}}}\!
=\!\left( \!\frac{P_{\text{
hyst}}}{P_{\text{eddy}}}\!\right)^{\!\!\!1/2}\!\!\!\!
=\left( \!\frac{3\mu ^{\prime \prime }}{\sigma
\omega \mu ^{\prime 2}t^{2}} \!\right)^{\!\!1/2}\!\!\!\!\!
=\sqrt{\frac{3}{2}}\frac{\delta _{\text{skin}}}{t
}\sqrt{\tan \delta _{\text{loss}}},
\label{ratio}
\ee
where we used the definitions of skin depth $\delta _{\text{skin}}=1/\sqrt{%
\pi \mu ^{\prime }\sigma f}$ and loss tangent $\tan \delta _{\text{loss}%
}=\mu ^{\prime \prime }/\mu ^{\prime }$. Therefore $\delta B_{\text{magn}}$ becomes relatively important when the skin depth is greater than $\sim t/\sqrt{\tan \delta_{\text{loss}}}$. This is equivalent to
$f\lesssim f_{\text{magn}}$ where
\begin{equation}\label{Eqmagn}
f_{\text{magn}}=3\tan\delta_{\text{loss}}/2\pi\mu_{r}\mu_{0}\sigma t^2.
\end{equation}
\subsection{Field noise equations}
In this section we list explicit formulas for the magnetic field
noise for shields of simple geometries shown in Fig. 1(c), namely an
infinite plate, infinite cylindrical shell, spherical shell and a
finite-length, closed cylindrical shell. From the considerations in the
previous sections, the on-axis mangetic field noise inside a cylindrically
symmetric, thin-walled shield can be calculated analytically from the
knowledge of $B_{\perp }(l)$. Calculation of $B_{\perp }$ is analogous to
that of an electric field on the inside surface of a conducting shell
induced by an on-axis electric dipole. Such calculation is most easily
performed by the method of an image in case of an infinite plate and a
sphere. For a cylinder, Smythe \cite{Smythe1968} gives a series expansion
solution that can be readily adopted for calculation of $B_{\perp }$.

\smallskip

{\it infinite plate}

The midplane of the plate is the $x$-$y$ plane, and the driving dipole $p\hat{z}$ is at $z=a$ on the $z$-axis. $l$ is measured from the origin. Due to the image effect of a high-permeability plate, $B_{\perp
}(l)$ is twice as large as the normal component of a dipolar field expected
in free space. Explicitly,
\[
B_{\perp }(l)=\frac{\mu _{0}p}{2\pi }\cdot \frac{-1+3\cos ^{2}\theta }{%
\left( a^{2}+l^{2}\right) ^{3/2}},
\]
where $\cos \theta =a/\sqrt{a^{2}+l^{2}}.$ This gives
\be
\delta B_{\text{curr}}=\frac{1}{\sqrt{6\pi }}\frac{\mu _{0}\sqrt{kT\sigma t}
}{a}.
\label{dB1}
\ee

{\it spherical shell}

For a driving dipole $p\hat{z}$ at the center of a sphere with
radius $a$, the image ``dipole'' consists of two ``monopoles'' $\pm 2ap/d^{2}
$ positioned at $z=\mp 2a^{2}/d$, in the limit $d\rightarrow 0.$ The
resulting surface normal field is
\[
B_{\perp }(l)=\frac{3\mu _{0}p}{4\pi a^{3}}\cos \frac{l}{a}
\]
where $l$ runs from the north pole to the south pole of the sphere, $0<l<\pi a$,
and
\be
\delta B_{\text{curr}}=\frac{1}{\sqrt{2\pi }}\frac{\mu _{0}\sqrt{kT\sigma t}
}{a}.
\label{dB2}
\ee

{\it infinite cylindrical shell}

Smythe \cite{Smythe1968} gives the electrostatic potential $V(\rho ,z)$ inside an
infinitely long conducting cylindrical tube, symmetric around the $z$-axis,
due to a point charge $q$ inside the tube. When $q$ is at the origin and the tube is grounded, it is
\[
V(\rho ,z)=\frac{q}{2\pi \epsilon _{0}a}\sum_{\alpha }e^{-\alpha \left|
z\right| /a}\frac{J_{0}(\alpha \rho /a)}{\alpha J_{1}^{2}(\alpha )}.
\]
where $\epsilon _{0}$ is the permittivity of vacuum, $J_{n}(x)$ is the
Bessel function of order $n$, and the summation is over the zeros of $J_{0}$%
; $J_{0}(\alpha )=0$. From this expression, the surface normal (radial)
magnetic field at $\rho =a$ due to a magnetic dipole $p\hat{z}$ at the
origin can be obtained as
\begin{eqnarray*}
B_{\perp }(z) &=&\mu _{0}\epsilon _{0}d\frac{\partial }{\partial z}\frac{%
\partial }{\partial \rho }V(\rho ,z)|_{\rho =a,\,\,qd=p} \\
&=&\text{sign}(z)\cdot \frac{\mu _{0}p}{2\pi a^{3}}\sum_{\alpha }\frac{%
\alpha e^{-\alpha \left| z\right| /a}}{J_{1}(\alpha )}.
\end{eqnarray*}
As a result,
\begin{eqnarray}
\delta B_{\text{curr}} &=&\frac{\mu _{0}\sqrt{kT\sigma t}}{a}\sqrt{\frac{2}{%
3\pi }G}, \label{dB3}\\
G &=&\int_{-\infty }^{\infty }dz^{\prime }\left( \sum_{\alpha }\frac{%
e^{-\alpha \left| z^{\prime }\right| }}{J_{1}(\alpha )}\right) ^{2}\approx
0.435.\nonumber
\end{eqnarray}
{\it closed cylindrical shell of finite length}

Ref.~[\onlinecite{Smythe1968}] also gives the electrostatic potential when the conducting cylinder
is closed, at, say,  $z=\pm L/2$, by conducting plates . For a
charge $q$ at $(\rho =0,\,z=z_{1})$, the potential at a point $\left( \rho<a
,z>z_{1}\right) $ is
\begin{eqnarray*}
&&V\left( \rho ,z\,;\,z_{1}\right)= \\&&\frac{q}{\pi \epsilon _{0}a}%
\sum_{\alpha }\frac{\sinh \alpha\!\left( \frac{L}{2a}+\frac{z_{1}}{a}\right)
\,\sinh \alpha\! \left( \frac{L}{2a}-\frac{z}{a}\right) }{\sinh \alpha \frac{L%
}{a}} \frac{J_{0}\left( \alpha \rho /a\right) }{\alpha J_{1}^{2}(\alpha
)}.
\end{eqnarray*}
If the conducting shell is replaced by a high-permeability magnetic shield
and a magnetic dipole $p\hat{z}$ replaces $q$, the normal magnetic field at
the top plate is
\begin{eqnarray*}
B_{\perp }^{top}&=&B_{z}\left( \rho ,z=L/2\right) \\&=&-\mu _{0}\epsilon _{0}d%
\frac{\partial }{\partial z}\frac{\partial }{\partial z_{1}}V\left( \rho
,z\,;\,z_{1}\right) |_{z=L/2,\,qd=p}.
\end{eqnarray*}
Similarly the normal field on the side wall at $z>z_1$ is
\begin{eqnarray*}
B_{\perp }^{side}&=&B_{\rho }\left( \rho =a,z\right) \\&=&-\mu _{0}\epsilon _{0}d%
\frac{\partial }{\partial \rho }\frac{\partial }{\partial z_{1}}V\left( \rho
,z\,;\,z_{1}\right) |_{\rho =a,\,qd=p}.
\end{eqnarray*}
For simplicity, in the following we consider only the case when $p\hat{z}$
is located at the origin, $z_{1}=0$, which gives the noise at the center of
the shield. Then by symmetry calculation of $%
\beta $ requires integral over only the upper half of the cylinder. The
integral path consists of two portions: the top plate where $l$ runs along
the line $(0<\rho <a,\,z=L/2)$ and the upper half of the side wall where $l$
runs along the line $(\rho =a,\,L/2>z>0)$. \newline
Explicitly,
\begin{eqnarray*}
&&\frac{1}{2}\beta =\int_{0}^{a}\frac{1}{\rho }d\rho \left[ \int_{0}^{\rho
}\!\!B_{\perp }^{top}\!\left( \rho ^{\prime }\right) \rho ^{\prime }d\rho ^{\prime
}\right] ^{2} \\
&&+\int_{0}^{L/2}\frac{1}{a}dz\left[ \int_{0}^{a}\!\!B_{\perp }^{top}\!\left( \rho
^{\prime }\right) \rho ^{\prime }d\rho ^{\prime }+\int_{z}^{L/2}\!\!B_{\perp
}^{side}\!\left( z^{\prime }\right) adz^{\prime }\right] ^{2}.
\end{eqnarray*}

The first term can be evaluated using Bessel function identities $%
\int_{0}^{u}u^{\prime }J_{0}\left( u^{\prime }\right) du^{\prime }=uJ_{1}(u)$
and $\int_{0}^{1}dx\,x\,J_{1}(\alpha x)J_{1}(\alpha ^{\prime }x)=\frac{1}{2}%
J_{1}^{2}(\alpha )\delta _{\alpha \alpha ^{\prime }}.$ This turns out to be $%
\left( \frac{\mu _{0}p}{2\pi }\right) ^{2}\frac{1}{2a^{2}}F_{1}\left(
L/a\right) $, where
\be
F_{1}=\sum_{\alpha }\frac{1}{\sinh ^{2}\frac{\alpha L}{2a}}\cdot \frac{1}{%
J_{1}^{2}(\alpha )}.
\label{F1}
\ee

The second term is more tedious, but can be reduced to $\left( \frac{\mu
_{0}p}{2\pi }\right) ^{2}\frac{1}{a^{2}}F_{2}\left( L/a\right) $ with \cite{footnote4}
\be
F_{2}=\int_{0}^{1/2}dx\frac{L}{a}\left[ \sum_{\alpha }\frac{1}{J_{1}\left(
\alpha \right) }\cdot \frac{\cosh \frac{\alpha Lx}{a}}{\sinh \frac{\alpha L}{%
2a}}\right] ^{2}.
\label{F2}
\ee

Finally the field noise is
\begin{eqnarray}
\delta B_{\text{curr}} &=&\frac{\mu _{0}\sqrt{kT\sigma t}}{a}\sqrt{\frac{2}{%
3\pi }G}, \label{dB4} \\
G &=&F_{1}(L/a)+2F_{2}(L/a).\nonumber
\end{eqnarray}
Numerical evaluation of the above equation shows that $G=0.657,\,0.460,%
\,0.438$ for aspect ratios $L/2a=1,\,1.5,\,2$, respectively. Thus the
noise from a closed cylindrical shield with aspect ratio of 2 already
approaches that of an infinitely long shield within 0.5\%.

\section{Comparison with noise from nonmagnetic conducting shells}

\begin{table*}[t]
\begin{tabular}{|c|c|c|}
\hline
\multirow{2}{*}{geometry} &
\multicolumn{2}{|c|}{field noise due to Johnson noise current}  \tabularnewline \cline{2-3}
& high-permeability & non-magnetic \\ \hline
infinite plate &

$\delta B_{\text{curr}}= {\displaystyle {1\over \sqrt{6\pi }} }{\displaystyle{\mu _{0}\sqrt{kT\sigma t}
\over a}}$
\rule{0pt}{5ex}\rule[-4ex]{0pt}{0pt}&
$\delta B=
{\displaystyle {1 \over \sqrt{8\pi }}}
{\displaystyle {   \mu _{0}\sqrt{kT\sigma t} \over a  }  }
$ \\ \hline
spherical shell &
$\delta B_{\text{curr}}= {\displaystyle {1 \over \sqrt{2\pi} } }{\displaystyle{\mu _{0}\sqrt{kT\sigma t}
\over a}}$
\rule{0pt}{5ex}\rule[-4ex]{0pt}{0pt}&
$\delta B=\sqrt{
{\displaystyle {2 \over 3\pi }}
}{\displaystyle {\mu _{0}\sqrt{kT\sigma t} \over a}}
$ \\ \hline
infinite cylindrical shell &
$\delta B_{\text{curr}} \!\!=\! \sqrt{ {\displaystyle{2G\over
3\pi} }  }\,{\displaystyle { \mu _{0}\sqrt{kT\sigma t} \over a}}$, $G \!\approx\!\! 0.435$
\rule{0pt}{5ex}\rule[-4ex]{0pt}{0pt}&
$\delta B=\sqrt{
{\displaystyle {3 \over 16}}%
}
{\displaystyle {\mu _{0}\sqrt{kT\sigma t} \over a}}
$ \\ \hline
finite, closed  &
\multirow{2}{*}{ Eqs.~(\ref{F1},\ref{F2},\ref{dB4}) } \rule{0pt}{5ex}\rule[-4ex]{0pt}{0pt}&
$\delta B=\sqrt{G}\,
{\displaystyle {\mu _{0}\sqrt{kT\sigma t} \over a}}
$ ,\\
cylindrical shell&
\rule{0pt}{2ex}\rule[-4ex]{0pt}{0pt}&
$G\!=\!
{\displaystyle {1 \over 8\pi }}
\left( { \displaystyle{ 3(L/2a)^5 + 5(L/2a)^3 + 2 \over
(L/2a)^2\left( 1+(L/2a)^{2}\right)^2 }} + 3\tan^{\!-1}\!
{\displaystyle {L \over 2a}}
\right) $ \\ \hline
\end{tabular}
\caption{\label{Tabcomp}
 Magnetic field noise from high-permeability and non-magnetic plate and
shells of conductivity $\sigma $. The geometries are shown in Fig. 1(c).
}
\end{table*}

An interesting question is how the magnetic field noise in a
high-permeability shield compares with that in a non-magnetic shell with the
same geometry and conductivity. As indicated in Ref.~[\onlinecite{Clem1987}], calculation of
low-frequency eddy current loss in an axially symmetric, nonmagnetic metal
driven by an axial dipole $\vec{p}=p\hat{z}\sin \omega t$ is relatively
simple, because the amplitude of the induced electric field is
proportional to the magnetostatic vector potential $A_{\phi }$ (in Coulomb
gauge) due to a dipole in vacuum.  For an axial dipole $p\hat{z}$ at the
origin
\[
A_{\phi }(\rho ,z)=\frac{\mu _{0}p}{4\pi }\frac{\rho }{\left( \rho
^{2}+z^{2}\right) ^{3/2}}
\]
and
\[
P_{\text{eddy}}=\frac{1}{2}\sigma \omega ^{2}\int_{V}A_{\phi }^{2}dv.
\]

Equations for the quasi-static field noises associated with this loss are
listed in Table \ref{Tabcomp} for the geometries considered in the previous section. It
is found that the current-induced noise inside a high-permeability shell is
in general not much different from that inside a non-magnetic shell. The
difference can be either positive (infinite plate) or negative (sphere and cylinder). Qualitatively, one can think of two competing effects, namely
self-shielding and image effects, due to the high permeability
of the material. In a long tube, the field generated by a noise current at
the end of the tube is self-shielded as it propagates inward. On the other
hand, the field generated by a current loop on the surface of an infinite
plate is amplified because of an image current adding field in the same
direction.

A dramatic illustration of the latter effect is found in the case of field
noise in between two infinite plates, with thickness $t$, separated by $L$.
When the plates are non-magnetic, the total quasi-static power loss induced
by an axial driving dipole half way between the plates is simply twice that
induced in a single plate. In the limit $\mu _{r}t/L\rightarrow \infty $,
however, it can be shown that the power loss and therefore the noise
logarithmically diverges. This is because the noise current in either plate
generates an infinite series of image currents, and when all the current
modes are considered their contributions do not converge. It is evident that
the noise from a high-permeability structure, even in the quasi-static
regime, cannot in general be obtained from the quadrature sum of the noise
from its individual parts.
\section{Frequency dependence}
\begin{table*}[t]

\begin{tabular}{|l|l|l|l|}
\hline
references & frequency dependence & material & method \\
\hline
Ref.~[\onlinecite{Sidles2003}], Eq.~(5) & $f^{0}\rightarrow f^{-1}\rightarrow f^{-3/4}$ &
non- or weakly magnetic slab & calculated \\
\hline
Ref.~[\onlinecite{Varpula1984}], Fig.~(6) & $f^{0}\rightarrow f^{-1}\rightarrow f^{-3/4}$ &
nonmagnetic slab & calculated \\
\hline
Ref.~[\onlinecite{Munger2005}], Fig.~(1) & $f^{0}\rightarrow f^{-1/4}\rightarrow
f^{-3/4}$ & high-permeability slab & calculated \\
\hline
Ref.~[\onlinecite{Roth1998}], Fig.~(2) & $f^{0}\rightarrow f^{-1}$ & nonmagnetic, thin sheet
& calculated \\
\hline
\multirow{2}{*}{Ref.~[20], Fig.~(2) }
& $f^{0}\rightarrow f^{-1/4}$ & mu-metal plate & measured\\
\cline{2-4}
& $f^{0}\rightarrow f^{-1}$ & copper plate & measured\\
\hline
\end{tabular}
\caption{\label{Tabfreq}Frequency dependence of magnetic field noise induced by Johnson
noise current in magnetic and non-magnetic metallic plates.}

\end{table*}

Here we consider how the noise $\delta B_{\text{curr}}$ considered in
Section III and IV rolls off at frequencies above the quasi-static regime.
Previous theoretical and experimental works on noise from
conducting plates and enclosures \cite{Nenonen1988} reported initial roll-off given by $\delta
B(f)\propto f^{-\gamma }$, where $\gamma \approx 1$ for non-magnetic metals
and $\gamma \approx 1/4$ for high-permeability metals. Below we provide
qualitative explanation of such dependences by considering a simple model.

Suppose we measure noise from a large, thin plate with conductivity $\sigma $
at a distance $a$ along the direction perpendicular to the plate. We assume that $\sigma$ is independent of frequency. The plate
has a thickness $t\ll a$ and a lateral dimension much larger than $a$. It is
reasonable to assume that the field noise mostly comes from
fluctuating currents flowing in a series of concentric rings directly below
the measurement point with radius on the order of $a$. Since these current
paths are connected in parallel, we can assume that in fact the noise comes from
current fluctuation in a single annular loop of mean radius $\approx a$ and width $%
\approx a$. The dc resistance of such a loop is $R_{0}= 2\pi /\sigma t$%
, which gives conventional Johnson noise current $\delta I=\sqrt{4kT/R_{0}}%
\approx \sqrt{(2/\pi )kT\sigma t}$. The magnetic field noise arising from
this current is indeed of the same order of magnitude as the noise
calculated in the previous sections.

At high frequencies this current is
suppressed in two ways. First, when $\delta _{\text{skin}}<t$, the
resistance increases by the skin depth effect to $R(f>f_{\text{skin}%
})\approx 2\pi /\sigma \delta _{\text{skin}}\propto f^{1/2}$. The threshold
frequency is
\begin{equation}\label{Eqskin}
f_{\text{skin}}=1/\pi \mu _{r}\mu _{0}\sigma t^{2}.
\end{equation}
Second, the self inductance $L$ of the loop suppresses $\delta I$ if $2\pi
fL>R(f)$. Therefore the current noise should in general be written as
open-loop voltage noise divided by total impedance,
\[
\delta I=\frac{\sqrt{4kTR(f)}}{|R(f)+i2\pi fL|}.
\]
where $R(f)$ includes the skin depth effect. If the condition $2\pi fL>R(f)$
is reached at a frequency $f_{\text{ind}}< f_{\text{skin}}$, such frequency is obtained from $2\pi f_{\text{ind}}L=R_{0}$, namely,
\begin{equation}\label{Eqind}
f_{\text{ind}}=1/\sigma tL= 1/C\mu _{0}\sigma ta,
\end{equation}
where $C$ is a constant of order unity. For a non-magnetic plate, $f_{\text{%
ind}}/f_{\text{skin}}$=$(\pi /C)\mu _{r}t/a\ll 1$ and inductive screening
indeed appears at a frequency far below that at which skin depth becomes
important. The initial roll-off of the noise then occurs at $f\gtrsim f_{%
\text{ind}},$ where the current noise scales with frequency as
\[
\delta I\approx \frac{\sqrt{4kTR_{0}}}{2\pi fL}\propto f^{-1},\,\,\,f_{\text{%
ind}}\lesssim f\lesssim f_{\text{skin}}.
\]
As $f$ further increases beyond $f_{\text{skin}}$, the scaling changes to
\[
\delta I\approx \frac{\sqrt{4kTR(f)}}{2\pi fL}\propto f^{-3/4},\,\,\,f_{%
\text{skin}}\lesssim f.
\]

On the other hand, for a high-permeability plate used for magnetic shields, skin
depth effect appears at a frequency far below that for inductive screening, $%
f_{\text{ind}}/f_{\text{skin}}\gg 1$. Therefore the initial roll-off is
expected to follow
\[
\delta I\approx \frac{\sqrt{4kTR(f)}}{R(f)}\propto f^{-1/4},\,\,\,f_{\text{%
skin}}\lesssim f\lesssim f_{\text{ind}}^{\prime }.
\]
The frequency $f_{\text{ind}}^{\prime }$ at which inductive screening
becomes important for a high-permeability plate is obtained from $2\pi f_{%
\text{ind}}^{\prime }L=R(f_{\text{ind}}^{\prime })=2\pi /\sigma \delta _{%
\text{skin}}$ which reduces to
\begin{equation}\label{Eqindp}
f_{\text{ind}}^{\prime }=(\pi /C^{2})\,\mu _{r}/\mu _{0}\sigma a^{2}.
\end{equation}
Beyond this frequency $\delta I\,$again scales as $f^{-3/4}$.

Table \ref{Tabfreq} summarizes the frequency dependence of the Johnson-current-induced
magnetic field noise reported in five references. It is found that our
simple model correctly predicts all the essential features of the frequency
dependences found in these works. For non-magnetic plates, the two threshold
frequencies Eq.~(\ref{Eqskin}) and Eq.~(\ref{Eqind}) agree, up to a numerical factor,
with those obtained in Ref.~[\onlinecite{Varpula1984}] \cite{footnote5} and Ref.~[\onlinecite{Sidles2003}] \cite{footnote6}. For high-permeability plates, Table (1) of Ref.~[\onlinecite{Munger2005}] also can be interpreted as giving the same threshold
frequencies between different regimes, Eq.~(\ref{Eqskin}) and Eq.~(\ref{Eqindp}),
obtained in this work \cite{footnote7}.

Finally, if we include the magnetization-fluctuation noise calculated in
Section III-B, the magnetic field noise from a high-permeability plate is expected to exhibit a rather complicated frequency dependence
\[
\delta B(f):f^{-1/2}\rightarrow f^{0}\rightarrow f^{-1/4}\rightarrow f^{-3/4}
\]
where the three threshold frequencies dividing different scaling regimes are
given by Eq.~(\ref{Eqmagn}), Eq.~(\ref{Eqskin}), and Eq.~(\ref{Eqindp}), in the increasing order.
\section{Noise reduction by differential measurement}
\begin{table*}[t]
\begin{tabular}{|c|c|p{2.5in}|c|}
\hline
\centering geometry & material &
\centering $\delta B_{\text{diff}}$ & $\delta B_{\text{diff}}/\delta B_{\text{single}}$
\tabularnewline \hline

\multirow{2}{*}{
\centering \includegraphics[width=0.8in]{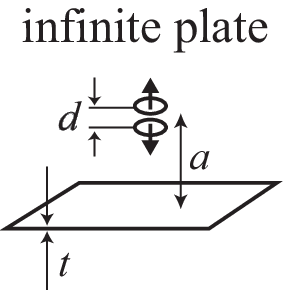}
}&
high-permeability \rule{0pt}{5ex}\rule[-4ex]{0pt}{0pt}
&
${\displaystyle {1 \over \sqrt{4\pi }}}
{\displaystyle {\mu _{0}\sqrt{kT\sigma t} \over a}}
\cdot {\displaystyle {d \over a}}$
&
\centering $ 1.22{\displaystyle{d \over a}}$
\tabularnewline \cline{2-4}
& non-magnetic \rule{0pt}{6ex}\rule[-4ex]{0pt}{0pt}
&
$\sqrt{
{\displaystyle {3 \over 16\pi }}}
{\displaystyle {\mu _{0}\sqrt{kT\sigma t} \over a}}
\cdot{\displaystyle {d \over a}}^*\footnote[0]{$^{*}$This agrees with Eq.~(43) of Ref.~[\onlinecite{Varpula1984}]}$
&
\centering $1.22{\displaystyle{d \over a}}$
\tabularnewline \hline
\multirow{3}{*}{
\includegraphics[width=1.3in]{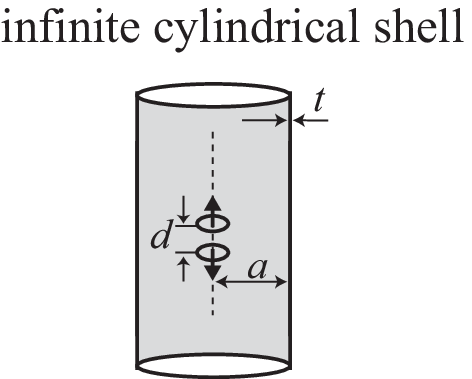}
}&
high-permeability \rule{0pt}{5ex}\rule[-2ex]{0pt}{0pt}
&
${\displaystyle { \sqrt{2 G\over 3\pi} } }
{\displaystyle {\mu _{0}\sqrt{kT\sigma t} \over a}}
\cdot{\displaystyle {d \over a}}$,
&
\centering $1.19{\displaystyle{ d \over a}}$
\tabularnewline
&&
$G = \displaystyle{
\int^{\infty}_{-\infty}dz^{\prime}
\left(\sum_{\alpha}\frac{\alpha e^{-\alpha \left|z^{\prime}\right|}}{J_{1}(\alpha) } \right)^2 \approx 0.618 }$
&
\\
\cline{2-4}
& non-magnetic \rule{0pt}{6ex}\rule[-4ex]{0pt}{0pt}
&
${\displaystyle {\sqrt{45\over 256}}}
{\displaystyle {\mu _{0}\sqrt{kT\sigma t} \over a}}
\cdot {\displaystyle {d \over a}}$
&
\centering $0.97{\displaystyle{ d\over a}}$
\tabularnewline \hline
\end{tabular}
\caption{\label{Tabdiff}
 Differential measurement noise from an infinite plate and a long cylindrical shell of conductivity $\sigma$. Arrows on two circular loops in each of the inset figures indicate the direction of the magnetic fields contributing to the measurement. $\delta B_{\text{single}}$ in the last column is the magnetic field noise in non-differential measurement taken from Table~\ref{Tabcomp}.
}
\end{table*}

A common technique to reduce the effect of magnetic field noise from a
distant source is to make a differential or gradiometric measurement. In the
first-order differential measurement, one measures $B_{
\text{diff}}(t)=B_{1}(t)-B_{2}(t)$, where $B_{1}$ and $B_{2}$ are the
magnetic fields at two points separated by a baseline $d$. The fluctuation
in this quantity $\delta B_{\text{diff}}(f)$ can be calculated following the
same principles described in Section II, with a single pickup coil
replaced by two coils connected in series so that the induced voltage is
proportional to $B_{\text{diff}}$. Reduction of noise from a distant source
now corresponds to reduction of power loss induced in the material when
driven by this ``gradiometric'' coil, which appears as a quadrupole, rather
than a dipole, seen from a distance $a\gg d$. If the two coils
connected in series are identical, each represented by an oscillating dipole of amplitude $p$, than the resulting power loss $P$ gives $\delta B_{\text{diff}}$ through
\[
\delta B_{\text{diff}}(f)=\frac{\sqrt{4kT}\sqrt{2P(f)}}{p\omega }.
\]
In the limit $a\gg d$, $P$ is proportional to the square of the driving quadrupole moment $%
p^{2}d^{2}$. From dimensional consideration, therefore, $\delta
B_{\text{diff}}$ scales as $(d/a)$.

Table \ref{Tabdiff} shows the results of analytical calculations of $\delta B_{\text{diff}}$ for an infinite plate and an infinitely long cylindrical shell. Only the white noise associated with the eddy current loss is considered. The noise is calculated for an axial differential measurement along the symmetry axis, in the limit where the baseline is much smaller than the shortest distance $a$ to the material. It is seen that in all cases the noise reduction factor is very nearly $d/a$.

\section{Conclusion}
We have used generalized Nyquist relation applied to electromagnetic power dissipation and magnetic field fluctuation to calculate magnetic field noise inside high-permeability magnetic shields. Analytical results for axially symmetric geometries show that the quasi-static field noise due to Johnson current noise in a metallic shell is slightly altered as the material gains high magnetic permeability. For magnetic shields with small electrical conductivity, $1/f$ noise from magnetization fluctuations becomes dominant over Johnson-current-induced noise below a threshold frequency proportional to its magnetic loss factor. Established numerical methods of finite-element analysis of electromagnetic power loss can be of great utility in calculating magnetic field noise spectrum from dissipative materials of complicated geometry. At relatively high frequencies, one could experimentally determine the power loss in dissipative materials using a pickup coil. This has an advantage that no prior knowledge of material parameters is necessary to predict the field noise. From Eq.~(\ref{Pf}) and (\ref{Bfromp}), it turns out that a 1 fT/Hz$^{1/2}$ noise at 1 kHz and at room temperature corresponds to an effective resistance of 10 m$\Omega$ in a 1000-turn driving coil of 5 cm diameter. This change in the resistive load is within the measurement range of modern impedance analyzers.

\begin{table*}[t]
\begin{tabular}{|c|c|c|c|c|}
\hline
\multirow{2}{*}{geometry} & \multicolumn{2}{|c|}{$\delta B_{\text{curr}}\,$%
[fT/Hz$^{1/2}$]} & \multicolumn{2}{|c|}{$\delta B_{\text{magn}}\,$[fT/Hz$%
^{1/2}$] at 1 Hz} \\ \cline{2-5}
& numerical & analytical & numerical & analytical \\ \hline
infinite plate & 3.63 & 3.68 & 2.01 & 2.07 \\
spherical shell & 6.38 & 6.38 & 3.57 & 3.59 \\
closed cylindrical shell &  &  &  &  \\
$L/2a=0.5$ & 12.4 & 12.2 & 6.93 & 6.87 \\
$L/2a=1.0$ & 6.01 & 5.97 & 3.33 & 3.36 \\
$L/2a=1.5$ & 5.04 & 4.99 & 2.77 & 2.81 \\
$L/2a=2.0$ & 4.92 & 4.87 & 2.70 & 2.74 \\ \hline
\end{tabular}
\caption{ \label{TableA1} Magnetic field noise calculated for mu-metal plate and enclosures
with $\sigma =1.6\times 10^{6}$ $\,\Omega ^{-1}$m$^{-1}$, $\mu _{r}=30,000$,
tan$\,\delta =0.04$. Geometrical parameters are $a=0.2$ m, $t=$ 1 mm,
referenced to Fig. 1(c). Column 3 is calculated from equations in column 2
of Table 1. Column 5 equals column 3 multiplied by 0.5628, from Eq.~(\ref{ratio}).
Numerical calculation for an infinite plate was obtained by extrapolation of
the results for finite-size plates.}
\end{table*}
\smallskip
\begin{table}
\begin{tabular}{|c|c|c|}
\hline
\multirow{2}{*}{geometry} & \multicolumn{2}{|c|}{$\delta B_{\text{curr}}$%
[fT/Hz$^{1/2}$]} \\ \cline{2-3}
& numerical & analytical \\ \hline
infinite plate & 15.4 & 15.5 \\
spherical shell & 35.7 & 35.9 \\
closed cylindrical shell &  &  \\
$L/2a=0.5$ & 45.0 & 44.9 \\
$L/2a=1.0$ & 34.1 & 34.2 \\
$L/2a=1.5$ & 33.6 & 33.7 \\
$L/2a=2.0$ & 33.6 & 33.7 \\ \hline
\end{tabular}
\caption{ \label{TableA2} Magnetic field noise from eddy current loss calculated in
aluminum, $\sigma $ $=3.8\times 10^{7}$ $\Omega ^{-1}$m$^{-1}$, $\mu _{r}=1$%
. The geometries are the same as in Table~\ref{TableA1}. }
\end{table}

    As reported earlier \cite{Allred2002}, we find that quasi-static Johnson current noise in magnetic shields is significantly higher than intrinsic noise of modern magnetometers.  Due to a small skin depth of high-permeability materials, however, the white noise range extends only to relatively low frequencies ($f_{\text{skin}}$ = 1$\sim$100 Hz), beyond which the noise rolls off as $f^{-1/4}$, until self-induction effect further brings down the noise. This indicates that usual room-temperature mu-metal shields may be used without adding significant noise if the signal is modulated at relatively high frequencies. At low frequencies, most sensitive experiments would require a low-loss nonconducting magnetic materials, such as certain ferrites, as the innermost layer of a multi-layer shield, or differential field measurement with a short baseline. In practice, a combination of these techniques should be implemented to suppress shield-contributed noise to an insignificant level.

\appendix
\section{Comparison with numerical calculations}
Here we compare magnetic field noises predicted by analytical expressions in
Table~\ref{Tabcomp} with those obtained from numerical calculations of power
loss for representative geometries. The calculation was performed by a
finite element analysis software (Maxwell 2D, Ansoft) which determined
electromagnetic fields in space on a mesh through
iterative solution of the Maxwell's equations. The driving dipole was
modeled as a small current loop on the symmetry axis. For $P_{\text{eddy}}$,
the current oscillated at $f$ = 0.01 Hz. For $P_{\text{hyst}}$, a
magnetostatic problem was solved with a static current in the coil, and
volume integral of $H^{2}$ in the material was calculated. Magnetic field
noises were then obtained by Eq.~(\ref{Bfromp}). The errors due to a non-zero radius of
the loop were insignificant within the accuracy of the numerical calculations
presented here.

Table~\ref{TableA1} shows magnetic field noises from high-permeability plate and shields.
The loss tangent assumed is for illustration purpose only. It is seen that in all cases considered here, numerical and analytical results differ by less than 3\%. The errors represent the accuracy of the assumptions made in magnetic
field calculations in Section III.

Table~\ref{TableA2} shows magnetic field noises from non-magnetic plate and shells. These
numbers can be used to estimate noises from non-magnetic, metallic enclosures
often used for radio-frequency shielding. The differences between
analytical and numerical calculations, less than 1\%, are consistent with the errors in the numerical calculations.

\section{Magnetic field noise from other metallic objects}

\begin{figure}
\includegraphics[width=8cm]{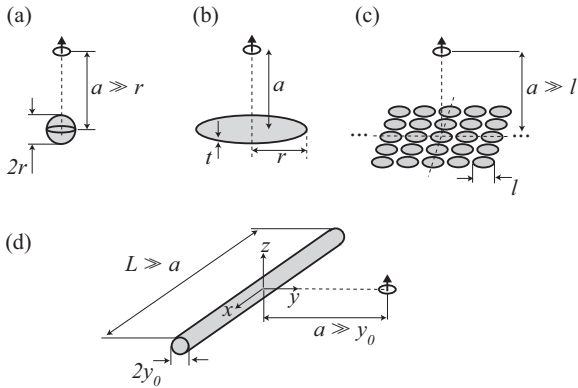}
\caption{ Definition of geometries used to calculate magnetic field noises from (a) a small solid sphere, (b) a thin disk with arbitrary diameter, (c) a planar, infinite thin film divided into a square array of small disks, and (d) a long, thin wire with a circular cross-section. In (c), the small disks are close-packed, but electrically isolated from each other. }
\label{figAppend}
\end{figure}
\begin{table*}
\begin{tabular}{|c|c|l|}
\hline
geometry & $\delta B$ & figure \\
\hline
small solid sphere \rule{0pt}{5ex}\rule[-3ex]{0pt}{0pt}
& $\sqrt{ {\displaystyle {4 \over 15\pi }} }\mu _{0}\sqrt{kT\sigma }\,r^{5/2}a^{-3}$ & (2a) \\
\hline
thin film disk \rule{0pt}{5ex}\rule[-3ex]{0pt}{0pt}
& $
{\displaystyle {1 \over \sqrt{8\pi }}}%
{\displaystyle {\mu _{0}\sqrt{kT\sigma t} \over a}}%
{\displaystyle {1 \over 1+a^{2}/r^{2}}}%
$ & (2b) \\
\hline
infinite square array of small disks \rule{0pt}{5ex}\rule[-3ex]{0pt}{0pt}
& $\sqrt{
{\displaystyle {3 \over 2048}}
}
{\displaystyle {\mu _{0}\sqrt{kT\sigma t\,}\,l \over a^{2}}}

$ & (2c) \\
\hline
circular cross-section wire \rule{0pt}{5ex}\rule[-3ex]{0pt}{0pt}
& $\sqrt{
{\displaystyle {3 \over 128}}
}\,\mu _{0}\sqrt{kT\sigma }\,y_{0}^{2}a^{-5/2}$ & (2d)\\
\hline
\end{tabular}
\caption{
\label{TableA3} Magnetic field noise due to Johnson noise current in metallic
objects with conductivity $\sigma $.
}
\end{table*}

For the purpose of future reference, here we list equations for magnetic
field noises resulting from Johnson noise currents in non-magnetic,
conducting objects with simple geometry. We only consider white noise in the
low frequency limit. Table~\ref{TableA3} lists equations for a small solid sphere, thin
planar films, and a long thin wire, as defined in Fig.~(\ref{figAppend}). In the context of an atomic vapor-cell
magnetometer, these objects can be associated with an alkali metal droplet,
low-emissivity conductive coatings on a glass, and a heating wire,
respectively. For problems with a cylindrical symmetry (Fig.~(\ref{figAppend}a,b) ), the eddy-current loss induced by a driving dipole $\vec{p}(t)=\vec{p}\sin \omega t$ was calculated by the method outlined in section IV. For others,  the eddy current density can be calculated from the equations
\begin{eqnarray*}
\nabla \times \vec{j} &=&\sigma \nabla \times \vec{E}=-i\sigma \omega \vec{B},
\\
\nabla \cdot \vec{j} &=&0
\end{eqnarray*}
with the boundary condition that the normal component of $\vec{j}$ is zero
on the surface (boundary) of the object. Here $\vec{B}$ is the amplitude of the
oscillating magnetic field generated by $\vec{p}(t)$ in free space. For a
thin film lying in the $x$-$y$ plane, $\nabla \times \vec{j}$ is along the $z
$ axis and therefore only $B_{z}$ contributes to the loss. The
two-dimensional current distribution $\vec{j}(x,y)=(u(x,y),v(x,y))$
then satisfies
\begin{eqnarray}
\frac{\partial u}{\partial y}-\frac{\partial v}{\partial x} &=&i\sigma
\omega B_{z} \label{jxy1}\\
\frac{\partial u}{\partial x}+\frac{\partial v}{\partial y} &=&0.\label{jxy2}
\end{eqnarray}

When the film is divided into small patches whose lateral dimensions are much smaller than the distance to the dipole, the current distribution in each patch can be calculated assuming a constant $B_{z}$ within the patch.

When the film is in the shape of a long, narrow strip, such as a long
straight wire patterned on an insulating substrate, the noise measured along
the $z$ axis on a point in the $x$-$y$ plane can be calculated by solving
Eqs.~(\ref{jxy1}, \ref{jxy2}) with the boundary condition $u(x\!=\!\pm
L/2,\,y)=0,\,v(x,y\!=\!\pm y_{0})=0$. Here the strip is assumed to occupy a
region $-L/2\leq x\leq L/2,$ $-y_{0}\leq y \leq y_{0}$ with $L\gg y_{0}$. The source term
is given by $B_{z}(x,y)=(\mu _{0}p/4\pi a^{3})(1+x^{2}/a^{2})^{-3/2}$,
assuming the noise is measured at $(0,a,0)$ and $a\gg y_{0}$. Eqs.~(\ref{jxy1}, \ref{jxy2}) are then satisfied by
\begin{eqnarray*}
u(x,y) &=&\frac{\sigma \omega \mu _{0}p}{\pi aL}\sum_{n}\cos k_{n}x\,\,\frac{%
\sinh k_{n}y}{\cosh k_{n}y_{0}}\,\,K_{1}(ak_{n}) \\
v(x,y) &=&\frac{\sigma \omega \mu _{0}p}{\pi aL}\sum_{n}\sin k_{n}x\left(
-1+\frac{\cosh k_{n}y}{\cosh k_{n}y_{0}}\right)K_{1}(ak_{n})
\end{eqnarray*}
where $K_{1}$ is the modified Bessel function of order one and $%
k_{n}=(2n-1)\pi /L,\,n=1,2,\cdots $. The power loss in a strip with thickness
$dt$, calculated in the limit $L/a\rightarrow \infty $, is
\begin{eqnarray*}
P(y_{0})dt &=&\frac{dt}{\sigma }\int \int (u^{2}+v^{2})\,dxdy \\
&=&\frac{dt}{64\pi }\,\sigma (\omega \mu _{0}p)^{2}\,y_{0}^{3}\,a^{-5}\left( 1-%
\frac{3}{8}\frac{y_{0}^{2}}{a^{2}}+\cdots \right) .
\end{eqnarray*}
For a wire with a circular cross-section, the integral of the above equation
over the profile (in the $y$-$z$ plane) of the wire gives the total loss
\[
P_{\text{wire}}=\int_{-y_{0}}^{y_{0}}P\!\left( \sqrt{%
y_{0}^{2}-t^{2}}\right) dt.
\]
The corresponding magnetic field noise, in the leading order
in $y_{0}/a$, is
\[
\delta B=\sqrt{\frac{3}{128}}\mu _{0}\sqrt{kT\sigma}\,y_{0}^{2}\,a^{-5/2}.
\]
For example, a long straight constantan wire with diameter 2$y_{0}$ = 1 mm
and $\sigma =$ 2$\times 10^{6}\,\Omega ^{-1}$m$^{-1}$ exhibits $\delta B=0.433$
fT/Hz$^{1/2}$ at room temperature when measured at $a$ = 1 cm in the
direction perpendicular to both the wire and the normal direction to the
wire.

\vspace{12pt}
\begin{acknowledgments}
The authors acknowledge helpful discussions with S. J. Smullin and T. W. Kornack, and their experimental work with a ferrite shield which inspired much of the present work. This research was supported by an Office of Naval Research MURI grant.
\end{acknowledgments}


\begin{thebibliography}{99}
\bibitem{footnote1}
For low-conductivity soft ferromagnets so-called excess or residual loss often dominates eddy current loss calculated for static bulk conductivity. In such cases we assume that the effect is included in appropriately re-defined $\sigma=\sigma(f)$.
\bibitem{footnote2}
This comes from the fact
that for a given shield size and shape, the reluctance of the portion of the
magnetic circuit that goes through the shield scales with the thickness and
the permeability as R$_{magn}\propto 1/(\mu _{r}t)$. Therefore qualitative
distribution of field lines around the shield should not change as $\mu _{r}$
and $t$ vary while keeping their product constant. The other length
scale relevant to the problem is $a$, which leads to the dimensionless
parameter specified.
\bibitem{footnote3}
Obviously this does not hold at the corners of a closed cylindrical shield. However, numerical finite-element calculations in Appendix A indicate that errors in noise due to these localized points are at most on the order of 1\%.
\bibitem{footnote7}
This is the case
after correcting the definition of $\zeta ^{2}$ by multiplying it with $z^{2}
$ in order to render it dimensionless as claimed.
\end{thebibliography}
\end{document}